
\def \({\left(}
\def \){\right)}
\def \n{\nabla}
\def \Tr{\hbox{Tr}}
\def \r{\rho}
\def \L{\Lambda}
\def \t{\tilde}
\def \O{\Omega}
\def \ss{\char'140} 
\def \pd{\partial}
\def \s{\sigma}
\def \b{\beta}
\def \a{\alpha}
\def \g{\gamma}
\def \d{\delta}

\def \l{\lambda}

\def \ph{\phi}

\def \Ph{\Phi}

\def \G{\Gamma}

\magnification=1200
\line{\hfil LTH282}
\line{\hfil April 1992}
\line{\hfil}
{\nopagenumbers
\font \bigbf=cmbx10 scaled \magstep 1

\tolerance=500
\line{\bigbf \hfil HIGHER ORDER CONFORMAL INVARIANCE OF
                   \hfil}
\line{\bigbf \hfil STRING BACKGROUNDS OBTAINED  \hfil}
\line{\bigbf \hfil BY O(d,d) TRANSFORMATIONS \hfil}
\vskip 1in
\line {\hfil \bf J. Panvel \hfil}
\vskip 12pt
\line {\hfil
\it DAMTP, University of Liverpool, Liverpool L69 3BX, UK \hfil}
\vskip .5in
\line {\hfil \bf Abstract \hfil}
\vskip 12pt
Proposals that $O(d,d)$ boosts of trivial backgrounds lead to non-trivial
conformally invariant backgrounds are checked to two loop order.
We find that conformal invariance can be achieved by adding simple
higher order corrections to the metric and dilaton.
\vfill
\eject}

\line{\hfil}
Recently the one-loop effective action for closed strings has been
shown to be invariant under an $O(d,d)$ transformation[1], generalizing
earlier results on scale-factor duality[2]. This leads to the
exciting possibility that new, highly non-trivial conformally invariant
backgrounds may be generated from simpler conformally invariant
backgrounds.
The invariance of the one loop action guarantees that such nontrivial
backgrounds will be conformally invariant up to one loop. However, the
conformal invariance of the transformed solution is not guaranteed at
higher order (though all orders arguments based on string field theory
have been advanced in ref.[3]).

To show the conformal invariance of the $\s$-model it is only required
to show the vanishing of $\bar\b^g_{i j}$ and $\bar\b^b_{i j}$,
which are functions
closely related to the metric and antisymmetric tensor $\b$-functions.
In this paper we take the non-trivial background
constructed by an $O(d,d)$ transformation in ref.[4] and calculate its
associated $\b$-functions to two-loop order. Although it is not found to be
conformally invariant by itself, the changes to the metric
and dilaton required at two loops are simple
enough to raise the hope that an exactly conformally invariant solution may be
explicitly constructed.

\line{\hfil}
The $O(d,d)$ invariance of the one-loop action was illustrated by
Meissner and Venezi\-ano in ref.[1]. Their argument shows
that starting from the usual low energy expansion of the one-loop
genus zero effective action for closed strings[5],
$$
  S^{(1)}=\int d^dx \sqrt{-g} \exp\{\phi\} \(\L -R
  -g^{ij}\pd_i \phi\pd_j\phi - {1\over12} H_{ijk} H^{ijk}\)
\eqno(1)$$
and assuming the metric and antisymmetric tensor fields $g_{ij}$
and $b_{ij}$ are functions of time
only, the action may be written in a form which is explicitly invariant under
$O(d,d)$ transformations.
$$
  S^{(1)}=\int dt \exp\{-\Ph\} \left(\L +(\dot\Ph )^2
          +{1\over 8}\Tr[\dot M\eta \dot  M\eta] \right)
\eqno(2)$$
General coordinate transformations together with $b_{ij} \rightarrow
b_{ij} + \pd_{[i}A_{j]}$ always allow $g$ and $b$ to be written as
$$
  g= \pmatrix{-1 & 0 \cr 0 & G(t) \cr} \qquad
  b= \pmatrix{ 0 & 0 \cr 0 & B(t) \cr}
\eqno(3)$$
The other symbols represent the $2d\times2d$ matrices[6]:

$$
  M= \pmatrix{ g^{-1} & -g^{-1}b \cr bg^{-1} & g -bg^{-1}b \cr} \qquad
  \eta = \pmatrix{ 0 & 1 \cr 1 & 0 \cr}\qquad
  \Phi = \ph - \ln \sqrt{\det g}
$$

It is now easy to see that a transformation that sends
$\Phi\rightarrow\Phi$,  $M\rightarrow \O M \O^T$, where $\O$
is defined such that $\O^T\eta\O =\eta$,
will leave the action unchanged, and that the matrices $\O$ are members of
the group $O(d,d)$.
This guarantees the one-loop conformal invariance of any background
created by an $O(d,d)$ transformation from another conformally invariant
background.

Taking a particularly simple exactly conformal background suggested as an
example by Gasperini, Maharana and Veneziano[4], we start with a flat
metric and zero dilaton and antisymmetric tensor:

$$
  g=\pmatrix {-1 & 0 & 0 \cr 0 & b^2 t^2 & 0 \cr 0 & 0 & 1} \qquad
  b=0 \qquad
  \ph= \hbox{ const (taken =0)}
\eqno(5)$$

New metrics may now be generated using an $O(d,d)$ transformation as above;
however it is found that backgrounds not related to the original in a trivial
fashion only arise when considering the $\ss\ss$boost" subgroup of $O(d,d)$
defined by the matrix
$$
  \O (\g) = \pmatrix{ 1+\t c & \t s   & \t c-1 & -\t s  \cr
                      -\t s  & 1-\t c & -\t s  & 1+\t c \cr
                      \t c-1 & \t s   & 1+\t c & -\t s  \cr
                      \t s   & 1+\t c & \t s   & 1-\t c \cr }
\eqno(6)$$
\line{\hfil}
where $\tilde c = \cosh (\g)$, $\tilde s = \sinh (\g)$ and $0<\g<\infty$.
For convenience and compactness we shall write all the following in terms
of $c=\cosh ({\g\over 2})$ and $s=\sinh({\g\over 2})$.

By calculating $M(\g)=\O^TM\O$ we may read off a new, boosted metric $g(\g)$,
a non-zero antisymmetric tensor $b(\g)$ and dilaton $\phi$:

$$
  g= \pmatrix{ -1 & 0 & 0 \cr
  0 & {cb^2t^2+s^2\over s^2b^2t^2+c^2} & {cs(b^2t^2+1)\over s^2b^2t^2+c^2}\cr
  0 & {cs(b^2t^2+1)\over s^2b^2t^2+c^2} & 1 \cr } \qquad
  b= \pmatrix{ 0 & 0 & 0 \cr
               0 & 0 & {cs(b^2t^2+1) \over s^2b^2t^2+c^2} \cr
               0 & {-cs(b^2t^2+1) \over s^2b^2t^2+c^2} & 0 \cr }
\eqno(7a)$$
\line{\hfil}
$$
  \ph = -\ln \left ( 1+{c^2 \over s^2}b^2t^2 \right )
\eqno(7b)$$

Now we shall consider the one-loop calculations of the metric and antisymmetric
tensor beta functions. As guaranteed by the $O(d,d)$ invariance of $S^{(1)}$
in Eqs(1),(2),
$\bar\b^{g(1)}_{i j}$ and $\bar\b^{b(1)}_{i j}$ vanish to this order
for the metric, antisymmetric tensor and dilaton given in Eqs(7).
However, to
ensure the cancellation of the two-loop $\b$-functions it is necessary to
explore the effect of higher order contributions to the metric and dilaton.
In fact, it turns out that the only changes necessary in Eqs(7) are to
replace $g_{00}$ by $-e^{\l(t)}$ and to allow for a higher order modification
to
the dilaton.

Replacing $g_{00}$ by $-e^{\l (t)}$ causes
remarkably few changes to the original Christoffel symbols. A new
$\G^0_{0 0}={1\over 2}\dot\l$ is introduced and $\G^0_{1 1}$, $\G^0_{1 2}$
are multiplied by $e^{-\l (t)}$, while the remainder are unchanged.
The new Riemann tensor components, denoted by a tilde, also have a simple form
in terms of the original components and $\l$:

$$
  \eqalign{
            \t R^1{}_{010} &= R^1{}_{010}+{1\over 2}\dot\l\G^1_{01} \cr
            \t R^0{}_{101} &=e^{-\l}\{R^0{}_{101}-{1\over2}\dot\l\G^0_{11}\}\cr
            \t R^0{}_{102} &=e^{-\l}\{R^0{}_{102}-{1\over2}\dot\l\G^2_{10}\}\cr
            \t R^2{}_{121} &= e^{-\l} R^2{}_{121} \cr }
\eqno(8)$$

Conformal invariance requires the functions $\bar\b^g_{ij}$ and
$\bar\b^b_{ij}$ to vanish. We find that this uniquely determines $\l$ and
the higher order contribution to the dilaton $\d\phi(t)$ up to two-loop order.
In terms of the actual renormalization group $\b$-functions
$\b^g_{ij}$ and $\b^b_{ij}$, potential diffeomorphisms $W_i$
and dilaton $\phi$ the functions $\bar\b^g_{i j}$, $\bar\b^b_{i j}$,
and $\bar\b^\ph$
(which is the central charge at the conformally invariant point) have
the following forms[7-9]:
$$ \eqalign{
   \bar\b^g_{ij} &= \b^g_{ij} + \n_i\n_j\ph +\n_{(i}W_{j)} \cr
   \bar\b^b_{ij} &= \b^b_{ij} + H^k{}_{ij}\n_k\ph +H^k{}_{ij}W_k\cr
   \bar\b^\phi &=\b^\phi-{1\over2}\n^k\ph\n_k\phi+{1\over2}\n^k\phi W_k\cr }
\eqno(9)$$
(We set the string tension $\a'=1$ throughout.)
At one-loop order $W_i$ is zero while at two loops it may take nonzero
values given by
$$
  W^{(2)}_j = \left(\mu+{1\over3}\right)\pd_i(H_{klm}H^{klm})
\eqno(10)$$
where $\mu$ corresponds to an ambiguity in the definition of the
dilaton. Here for convenience we choose $\mu = -{1\over3}$ and hence
$W_i =0$; however any choice of $\mu$ would yield equivalent results.

The vanishing of the metric and antisymmetric
tensor $\bar\b$-functions guarantee that $\bar\b^\phi$ is a constant[9-11]
(in our case it is found to be zero) so we
shall first evaluate $\bar\b^g_{ij}$ and $\bar\b^b_{ij}$ to one loop order
using
the metric, antisymmetric tensor and dilaton together with the higher order
corrections, $g_{00}\rightarrow -e^{\l(t)}$ and $\ph\rightarrow\ph+\d\ph$. The
one-loop $\bar\b$ functions in Eqs(9) are given by[12-14]:

$$ \eqalign{
   \bar\b^{g(1)}_{ij} &= R_{ij} -H_{ikl}H_j{}^{kl} + \a'\n_i\n_j\ph\cr
   \bar\b^{b(1)}_{ij} &= -\n^kH_{kij} + \a'H^k{}_{ij}\n_k\phi \cr }
\eqno(11)$$

Due to the $O(d,d)$ invariance of $S^{(1)}$, when we substitute the explicit
forms of the background fields into Eq.(11) all 1st order terms
cancel; however the corrections produce higher order terms proportional
to $\l$ or $\d\phi$ which can be arranged to cancel the nonzero functions
$\bar\b^{g(2)}_{i j}$ and $\bar\b^{b(2)}_{i j}$.

It is found to 2nd order (ie taking $e^{-\l}=1$ and
neglecting all products of $\l$ and $\d\phi$) using Eq.(8) that the
$\bar\b$-functions in Eq.(11) take the explicit form:

$$ \eqalign {
   \bar\b^{g(1)}_{00} &=   {1\over2}\dot\l(\G^1_{10}+\G^2_{20}-\dot\ph)
-\d\ddot\ph = {1\over 2t}\dot\l +\d\ddot\ph \cr
   \bar\b^{g(1)}_{11} &= -\G^0_{11}\({1\over2}\dot\l +\d\dot\ph\)e^{-\l}\cr
   \bar\b^{g(1)}_{12} &= -\G^0_{12}\({1\over2}\dot\l +\d\dot\ph\)e^{-\l}\cr
   \bar\b^{g(1)}_{22} &= 0 \cr }
\eqno(12)$$

It is interesting to note that neither the proposed change in $g$ nor
$\phi$ may produce any cancellation of a non-zero $\b^{g(2)}_{2 2}$ term.
We shall see that this corresponds to fixing the choice of renormalization
scheme used to evaluate $\b^{g(2)}_{i j}$.
Similarly the higher order corrections to $g$ and $\phi$ give rise to
non-zero components in the one-loop antisymmetric tensor $\bar\b$-function
which
provides the cancellation required by $\bar\b^{b(2)}_{i j}$.

$$
  \bar\b^{b(1)}_{12}= \G^0_{12}\left({1\over2}\dot\l+\d\dot\ph\right)e^{-\l}
\eqno(13)
$$

The forms of the two loop metric and antisymmetric tensor $\b$-functions
are considerably more complex and their calculation was greatly
simplified by use of programs written in REDUCE, while analytic checks
were made on the consistency of the results.
At two loop order the metric and antisymmetric tensor $\b$ functions may
be written[12,15]:
$$ \eqalign{
  \b^{g(2)}_{ij}(\l) = &{1\over2}R_{iklm}R_j{}^{klm} +{1\over6}\n_iH_{klm}
  \n_jH^{klm}\cr &+\left ( {1\over2}-\r \right ) \n_kH_{ilm}\n^kH_j{}^{lm}
  +\r R^k{}_{ijl}H_{kmn}H^{lmn} \cr &+(6-4\r )R^k{}_{lmi}H_{jnk}H^{lmn}
  +2(\r-1)R_{klmn}H_i{}^{km}H_j{}^{ln} \cr
  &+H_{klm}H_n{}^{lm}H^k{}_{ip}H_j{}^{pn}+H_{ikl}H_{jmn}H_p{}^{km}H^{pln} \cr
  \b^{b(2)}_{ij}(\l) = &-2R_{klmi}\n^lH_j{}^{km} +\n_k\left ( H_{ilm}H_j{}^{ln}
  \right ) H_n{}^{km} \cr &-2(\r-1)\n_k\left ( H_n{}^{lm}H_{lmi} \right )
  H_j{}^{nk}+\r\n_kH_{ijn}H^k{}_{lm}H^{nlm} \cr }
\eqno(14)$$
where $\r$ parametrises renormalization scheme ambiguity.

It is only required to evaluate these using the original metric,
antisymmetric tensor and dilaton in Eqs(7) as the higher order corrections
would produce terms of order $\geq 3$.

It must be shown that there exists some choice of $\l$ and $\d\phi$ for
which the sums $\bar\b^{g(1)}_{i j} + \bar\b^{g(2)}_{i j}$ and
$\bar\b^{b(1)}_{i j} +\bar\b^{b(2)}_{i j}$ vanish.
Since no change can be made to $\b^{g(2)}_{2 2}$ by $\b^{g(1)}_{11}$, it is
a prerequisite
that $\b^{g(2)}_{22}$ must itself be zero and indeed it is found that
this term is proportional to $(\r-1)$.

$$
  \b^{g(2)}_{22} = -{16c^2s^4b^4(\r-1)\over(s^2b^2t^2+c^2)^4}(s^2b^2t^2-c^2)
\eqno(15)$$

This fixes our choice of $\r$, removing the renormalization scheme
ambiguity present in the calculation of the two loop functions. Choosing
$\r=1$  allows the remaining components of $\b^{g(2)}_{i j}$ and
$\b^{b(2)}_{i j}$ to be written in a compact form:

$$ \eqalign{
   \b^{g(2)}_{00} &= -{4s^6b^6t^2\over(s^2b^2t^2+c^2)^3} \cr
   \b^{g(2)}_{11} &= -\G^0_{11}{4s^6b^6t^3\over(s^2b^2t^2+c^2)^3}\cr
   \b^{g(2)}_{12} &= -\G^0_{12}{4s^6b^6t^3\over(s^2b^2t^2+c^2)^3}\cr
   \b^{b(2)}_{12} &= \G^0_{12}{4s^6b^6t^3\over(s^2b^2t^2+c^2)^3}\cr
 }
\eqno(16)
$$

All other components are either identically zero term by term or are
trivially related to the above by symmetry.
Comparison of Eqs(12),(13),(16) shows a remarkable uniformity amongst the
two-loop components $\bar\b^{g(2)}_{11}$, $\bar\b^{g(2)}_{12}$,
$\bar\b^{b(2)}_{12}$, and the corresponding $\bar\b^{g(1)}_{ij}$,
$\bar\b^{b(1)}_{ij}$. As a consequence $\bar\b^{g(1)}_{ij}+\bar\b^{g(2)}_{ij}$
and $\bar\b^{b(1)}_{ij}+\bar\b^{b(2)}_{ij}$ can be arranged to vanish by
imposing only two constraints:

$$ \eqalign {
  {1\over2}\dot\l +t\d\ddot\ph &= {4s^6b^6t^3\over (s^2b^2t^2+c^2)^3} \cr
  {1\over2}\dot\l +\d\dot\ph   &= -{4s^6b^6t^3\over (s^2b^2t^2+c^2)^3} \cr }
\eqno(17)$$

Solving for a second order ODE in $\d\phi$ immediately gives $\d\dot\phi$
and hence $\dot{\l}$. These may
again be integrated analytically giving $\d\phi$ and $\l$ uniquely up to
constants.
The results are as follows:

$$ \eqalign {
  \d\dot\ph &= {-2s^4b^4t\over (s^2b^2t^2+c^2)^2}\cr
  \dot\l      &=-{4s^4b^4t(s^2b^2t^2-c^2)\over(s^2b^2t^2+c^2)^3} \cr }
   \eqalign {
  \d\ph       &= {s^2b^2 \over (s^2b^2t^2+c^2)}\cr
  \l          &= {2(s^4b^4t^2)\over(s^2b^2t^2+c^2)^2} \cr }
\eqno(18)$$

Although the dilaton $\bar\b$-function is now guaranteed to be a
constant it is a useful check on consistency to be
sure that this is indeed so. It is also a worthwhile exercise as the
value of the dilaton $\bar\b$-function relates to the conformal charge
contributed by the three fields $(t,x_1,x_2)$ considered.
The general form of the dilaton $\b$-function at one and two loops is given
by[7,9,11,16]:
$$ \eqalign{
  \b^{\phi(1)} = &\quad{1\over3}H_{klm}H^{klm} +{1\over2}\n^2\ph \cr
  \b^{\ph(2)}  = &-{1\over8}R_{klmn}R^{klmn} +{1\over2}(5+12\mu)
  R_{klmn}H_p{}^{ln}H^{pkm} \cr  &-\({1\over6}+\mu\)\n_kH_{lmn}\n^kH^{lmn}
  +\({1\over4}+\r +3\mu\)H_{kmn}H_l{}^{mn}H^{kpq}H^l{}_{pq} \cr
  &-{5\over12}H_{klm}H^k{}_{np}H_q{}^{ln}H^{qmp} -{1\over2}\n_k\n_l\ph
  H^{kmn}H^l{}_{mn} \cr }
\eqno(19)$$

As discussed earlier we take for convenience $\mu=-{1\over3}$ (so that
$W^{(2)}_i =0$ ) and we must also take $\r=1$ to ensure that $\b^{g(2)}_{22}$
vanishes. With these values for $\mu$ and $\r$, and with the background of
Eqs(7) together with the corrections to $g_{00}$ and $\ph$, we find from
Eqs(7),(9),(19), that the $\bar\b^{\ph(1)}$ and $\bar\b^{\ph(2)}$ reduce to:
$$ \eqalign {
  \bar\b^{\ph(1)}
&=-{\d\ddot\ph\over2e^\l}-{\d\dot\ph(3s^2b^2t^2+c^2)+s^2b^2t^2
                    \dot\l\}\over 2e^\l(s^2b^2t^2 +c^2)t} \cr
  \bar\b^{\ph(2)} &= -{s^4b^4\(2s^4b^4t^4+2c^4\)\over(s^2b^2t^2+c^2)^4} \cr }
\eqno(20)
$$
so that with $\d\dot\ph$ and $\dot\l$ as given by Eq.(18), $\bar\b^{\ph(1)}+
\bar\b^{\ph(2)}$ vanishes to this order. Hence the contribution to the central
charge from the $(t,x_1,x_2)$ sector of the theory is zero, and this seems
likely to persist to all orders. (Of course additional free fields can be
added to ensure the correct critical value of the central charge.)

\line{\hfil}
We have shown that although the one-loop beta functions for a boosted
metric vanish, at two-loops simple corrections to the metric and dilaton are
required to preserve conformal invariance. This is analogous to results
found by Tseytlin[17] in the context of conventional duality.

Sen[3] argues, using string field theory that the $O(d,d)$ invariance
should persist to all orders in perturbation theory. He suggests that
the transformed metric, dilaton and antisymmetric tensor might require
corrections at a higher order, which he speculates might be expressible
in terms of covariant quantities, e.g. $\d g_{i j} \propto R_{i j}$.
The corrections we have found do not appear to be of this form, which
raises doubts as to whether the $O(d,d)$ invariance is exactly preserved
at higher orders. Nevertheless the corrections we have found at two loops
are sufficiently simple to suggest that there may be an explicit modified
version of
the transformed solution which is conformally invariant to all orders in
perturbation theory.
In another recent paper[18] Sen shows how to construct solutions with
torsion by acting with $O(d,d)$ transformations on the 2-$d$ string
black hole solution of Witten[19]. Based on the results we have found
here, we speculate that again simple modifications might suffice to
render such solutions valid at two loops and higher.

\line{\hfil}
\line{\bf Acknowledgements \hfil}
\line{\hfil}
\noindent
J.P. thanks the SERC for financial support and Ian Jack for
advice and discussions.

\line{\hfil}
\line{\bf References  \hfil}
\line{\hfil}

\item {1. } K.A. Meissner and G. Veneziano, Phys. Lett. B267 (1991) 33.
\item {2. } G. Veneziano, Phys. Lett. B265 (1991) 287.
\item {3. } A. Sen, Phys. Lett. B271 (1991) 295.
\item {4. } M. Gasperini, J. Maharana and G. Veneziano, Phys. Lett.
            B272 (1991) 277.
\item {5. } C. G. Callan, D. Friedan, E. J. Martinec and M. J. Perry,
            Nucl. Phys. B262 (1985) 593.
\item {6. } A. Giveon, E. Rabinovici and G. Veneziano,
	    Nucl. Phys. B322 (1989) 167;
\item {   } A. Shapere and F. Wilczek, Nucl. Phys. B320 (1988) 669.
\item {7. } A. A. Tseytlin, Phys. Lett. B178 (1986) 34.
\item {8. } G. M. Shore, Nucl. Phys. B286 (1987) 349.
\item {9. } A. A. Tseytlin, Nucl. Phys. B294 (1987) 383.
\item {10.} G. Curci and G. Paffuti, Nucl. Phys. B286 (1987) 399.
\item {11.} H. Osborn, Ann. Phys. (N.Y.) 200 (1990) 1.
\item {12.} D. Friedan, Phys. Rev. Lett. 45 (1980) 1057;
            Ann. Phys. (N.Y.) 163 (1985) 318.
\item {13.} A. A. Tseytlin, Nucl. Phys. B276 (1986) 391.
\item {14.} T. L. Curtright and C. K. Zachos, Phys. Rev. Lett. 53 (1984) 1799;
\item {   } E. Braaten, T. L. Curtright and C. K. Zachos,
  	    Nucl. Phys. B260 (1985) 630.
\item {15.} C. M. Hull and P. K. Townsend, Phys. Lett. B191 (1987) 115;
\item {   } R. R. Metsaev and A. A. Tseytlin, Phys. Lett. B191 (1987) 363;
\item {   } D. Zanon, Phys. Lett. B191 (1987) 354;
\item {   } D. R. T. Jones, Phys. Lett. B192 (1987) 391;
\item {   } I. Jack and D. R. T. Jones, Phys. Lett. B193 (1987) 449.
\item {16.} I. Jack and D. R. T. Jones, Phys. Lett. B200 (1988) 453;
\item {   } C. M. Hull and P. K. Townsend, Nucl. Phys. B301 (1988) 197.
\item {17.} A. A. Tseytlin, Mod. Phys. Lett A6 (1991) 1721.
\item {18.} A. Sen, Phys. Lett. B274 (1992) 34.
\item {19.} E. Witten, Phys. Rev. D44 (1991) 314.

\end